# Additional Evidence for the Surface Origin of the Peculiar Angular-Dependent Magnetoresistance Oscillations Discovered in a Topological Insulator $Bi_{1-x}Sb_x$

**A A Taskin, Kouji Segawa and Yoichi Ando**[1]

Institute of Scientific and Industrial Research, Osaka University
Ibaraki, Osaka 567-0047, Japan

E-mail: y_ando@sanken.osaka-u.ac.jp

**Abstract**. We present detailed data on the unusual angular-dependent magnetoresistance oscillation phenomenon recently discovered in a topological insulator $Bi_{0.91}Sb_{0.09}$. Direct comparison of the data taken before and after etching the sample surface gives compelling evidence that this phenomenon is essentially originating from a surface state. The symmetry of the oscillations suggests that it probably comes from the (111) plane, and obviously a new mechanism, such as a coupling between the surface and the bulk states, is responsible for this intriguing phenomenon in topological insulators.

## 1. Introduction

Topological insulators [1-3] are an emerging class of materials that host a new quantum-mechanical state of matter [4-7] where an insulating bulk state supports an intrinsically metallic surface state that is "topologically protected", meaning that it is stable against any disorder that does not break time-reversal symmetry. Recently, angle-resolved photoemission spectroscopy (ARPES) studies on a cleaved trigonal surface of $Bi_{1-x}Sb_x$ have revealed that the energy dispersions of its surface states possess the distinctive character to qualify this material as a topological insulator [8-10]. However, to directly probe the unique properties of the surface states and to elucidate whether they could really be exploited on a macroscopic level, transport and magnetic studies of high-quality single crystals are indispensable. Unfortunately, in "real-life" samples of topological insulators available today, there is always some bulk conductivity due to residual carriers, and separating the contributions from two-dimensional (2D) and three-dimensional (3D) states turns out to be challenging [11-13].

In this context, we have recently succeeded in observing both the de Haas-van Alphen (dHvA) oscillations [14] and the Shubnikov-de Haas (SdH) oscillations [15] in high-quality bulk single crystals of $Bi_{1-x}Sb_x$ alloy in the "insulating" regime ($0.07 \leq x \leq 0.22$), which is the first material to be known as a 3D topological insulator. These observations became possible by growing highly pure and homogeneous single crystals of this alloy and achieving the bulk carrier density of the order of $10^{16}$ $cm^{-3}$. The dHvA and SdH oscillations signified a previously-unknown Fermi surface (FS) with a clear 2D character that coexists with a 3D bulk FS [14,15]. Since $Bi_{1-x}Sb_x$ is a 3D material, the observed 2D FS is naturally assigned to the "surface", which could be internal surfaces such as twin boundaries.

---
[1] To whom any correspondence should be addressed.

Furthermore, we have extended our measurements to the angular dependence of the magnetoresistance (MR), which was successfully applied to the studies of quasi-2D organic conductors in the 1980s and lead to the discovery of the celebrated angular-dependent magnetoresistance oscillations (AMRO) [16-18]. Intriguingly, in our angular-dependent MR studies, we found oscillatory angular dependences in both the resistivity $\rho_{xx}$ and the Hall resistivity $\rho_{yx}$ [15]. The oscillations observed at lower fields were elucidated to be essentially a manifestation of the SdH oscillations, whereas the ones observed at higher fields were obviously of novel origin [15]. We proposed that the latter originates from the topological surface state on the cleaved (111) surface of $Bi_{1-x}Sb_x$, where the coupling between the surface and bulk states are probably playing a key role in the peculiar MR oscillation phenomenon [15].

In this paper, we present additional evidence to demonstrate that the "high-field" oscillations are most likely due to the exposed surface, by directly comparing the oscillations before and after the sample surface was chemically etched.

## 2. Experimental

High-quality $Bi_{0.91}Sb_{0.09}$ single crystals were grown from a stoichiometric mixture of high-purity (6N) Bi and Sb elements by a zone melting method. The resistivity was measured by a standard four-probe method on a rectangular sample, where the current was directed along the $C_1$ axis. Continuous rotations of the sample in constant magnetic fields were used to measure the angular dependence of the MR within the trigonal-binary ($C_3$-$C_2$) crystallographic plane. Magnetic fields up to 16 T were applied using a dc superconducting magnet in a $^4$He cryostat. All the data shown here were taken at the lowest temperature of 1.5 K. To obtain a fresh surface on the $Bi_{1-x}Sb_x$ single crystals, we applied the following chemical etching procedure:

1) Keep for several minutes in the 1:4:9 mixture of $HNO_3$, $CH_3COOH$, and $H_2O$.
2) Rinse in distilled water.
3) Keep for several minutes in the diluted (50 mol-%) HCl.
4) Rinse in distilled water and dry.

## 3. Results

*3.1. Angular-Dependent MR Oscillations*

Figures 1(a) and 1(b) show the angular dependences of the transverse MR and the Hall resistivity,

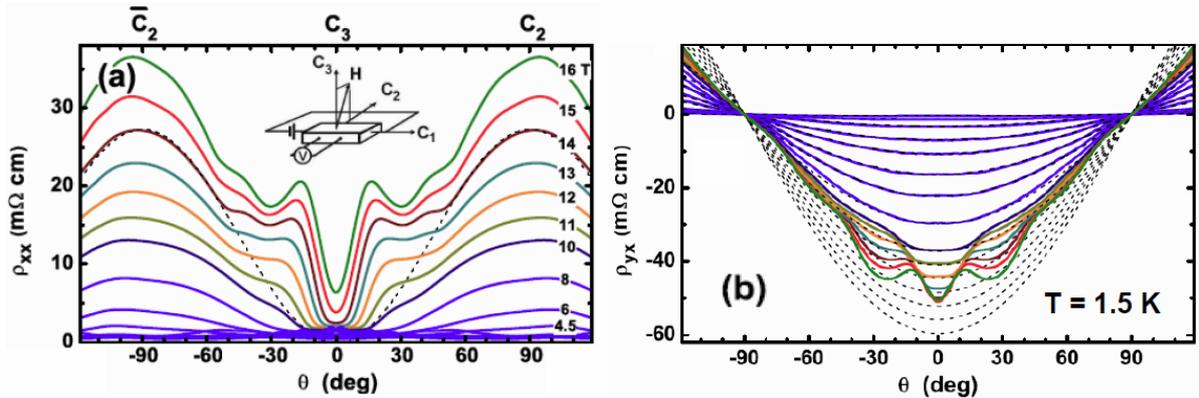

Figure 1. (a) Angular dependences of $\rho_{xx}$ measured in the trigonal-binary ($C_3$-$C_2$) plane in constant magnetic fields. The dashed line is the MR background ~ $(1-\cos^2\theta)$ for $B = 14$ T. Inset depicts the measurement geometry, where $\theta = 0°$ corresponds to the $C_3$ axis. (b) Angular dependences of $\rho_{yx}$ measured in the same conditions as in (a). The dashed lines are the expected angular dependences of the Hall effect plotted for all magnetic fields.

respectively, measured in magnetic fields rotated within the trigonal-binary ($C_3$-$C_2$) plane perpendicular to the current. The magnetic-field strength was kept constant during each rotation. As can be seen in Figs. 1(a) and 1(b), both $\rho_{xx}$ and $\rho_{yx}$ present an oscillatory behavior as a function of the rotation angle $\theta$. As we have elaborated in our previous paper [15], two different types of oscillations can be distinguished: The first type [though not very clearly seen in Fig. 1] consists of oscillations appearing at lower fields, while the second one becomes prominent at higher fields ($B \geq 10$ T). The former "low-field" oscillations were demonstrated [15] to be essentially due to the SdH oscillations originating from the 2D state residing on the (2,-1,-1) plane, which was unambiguously observed in the $\rho_{xx}(B)$ data for magnetic-field sweeps in fixed magnetic-field directions.

The focus of the present paper is the "high-field" oscillations, which develop on a smooth field-dependent background coming from the anisotropy of $\rho_{xx}(B)$ along the different axes. An example of the fitting of the background for $B = 14$ T is shown by the dashed line in Fig. 1(a). Because of a large MR background in strong magnetic fields, only largest peaks in $\rho_{xx}(\theta)$ can be clearly seen in the raw data [Fig. 1(a)].

The $\rho_{yx}(\theta)$ data shown in Fig. 1(b) also present pronounced angular-dependent oscillations at high fields. The "background" for $\rho_{yx}(\theta)$ is simply the angular dependence of the Hall effect, $R_H B \cos\theta$, where $R_H$ is the Hall coefficient. As can be clearly seen in Fig. 1(b), low-filed $\rho_{yx}(\theta)$ data follow this expected angular dependence very closely (we use $R_H = -37$ cm$^3$/C, obtained from the Hall measurements), and the large deviation from this simple behavior is observed only in magnetic fields above 10 T. Figure 2 shows "pure" oscillations in $\Delta\rho_{yx}(\theta)$ after subtracting the $R_H B \cos\theta$ contribution from $\rho_{yx}(\theta)$. One can clearly see a set of peaks, which are marked by short vertical ticks in Fig. 2. They are symmetric with respect to the $C_3$ axis and show a rather complicated magnetic-field dependence.

### 3.2. Surface-Condition Dependence

The data shown in Figs. 1 and 2 were taken on a refreshed surface immediately after chemically etching the sample. To investigate the effect of etching, we had conducted the same angular-dependent MR measurements on the same sample before etching. Incidentally, this sample was characterized 7 months before those measurements, and it was kept in a desiccator, exposing its surface to dry air for 7 months. Our intention was to see how such an "old" sample behaves. Intriguingly, the angular-dependent MR oscillations were smeared, as shown in Fig. 3 for $\rho_{xx}$. Figure 4 presents a direct comparison of the oscillations in $\rho_{xx}$ and $\rho_{yx}$ before and after etching. It is evident that the refreshed surface yields more pronounced oscillations, which gives direct evidence that the peculiar angular-dependent oscillations are essentially originating from a surface state.

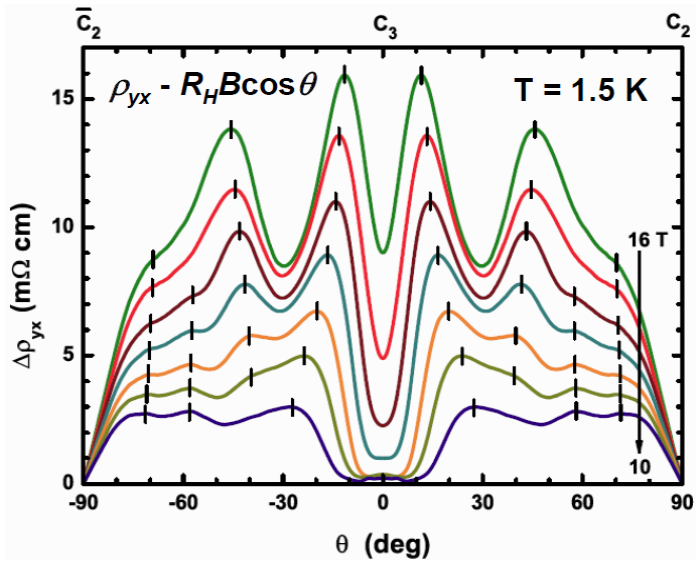

Figure 2. "High-field" angular-dependent oscillations in $\Delta\rho_{yx}(\theta)$ obtained by subtracting the expected $R_H B \cos\theta$ behavior from the $\rho_{yx}$ data shown in Fig. 1(b). Ticks mark the positions of distinguishable peaks.

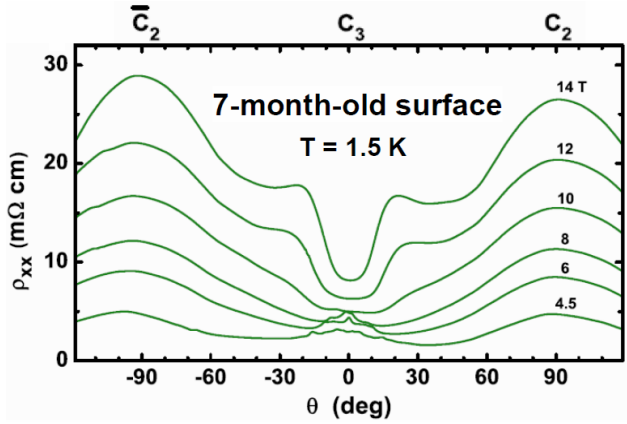

Figure 3. Angular dependences of $\rho_{xx}$ measured in the trigonal-binary ($C_3$-$C_2$) plane on the 7-month-old surface of the same $Bi_{0.91}Sb_{0.09}$ sample before etching. Note that the angular-dependent MR oscillations are smeared compared to those shown in Fig. 1(a).

## 4. Discussions

In the "high-field" oscillations, an important feature is that the amplitude of the peaks weakens as the magnetic field is rotated away from the $C_3$ axis, which is somewhat reminiscent of the behavior of the ordinary AMRO in quasi-2D systems if the conduction planes lie perpendicular to the $C_3$ axis [16-18]. Thus, it is probable that the "high-field" oscillations are coming from the (111) plane (which is perpendicular to the $C_3$ axis), where surface states are seen in photoemission [8-10] and tunnelling [19] experiments. Another distinguishable feature of the "high-field" oscillations is that they survive up to rather high temperatures [15]. For example, even at 40 K there are still visible traces of oscillations while the SdH oscillations are already gone at this temperature [15], which is reminiscent of the behaviour of the ordinary AMRO in quasi-2D systems. In spite of these similarities to the quasi-2D AMRO, the peak positions of the "high-field" oscillations apparently shift with the magnetic field, which is not expected for the ordinary AMRO. Moreover, the existence of a finite coupling between conduction planes is essential for the quasi-2D AMRO [20], but there is no such inter-plane coupling for the surface states as long as the crystal is thick enough. Therefore, the observed "high-field" angular oscillations are a new phenomenon apparently specific to topological insulators.

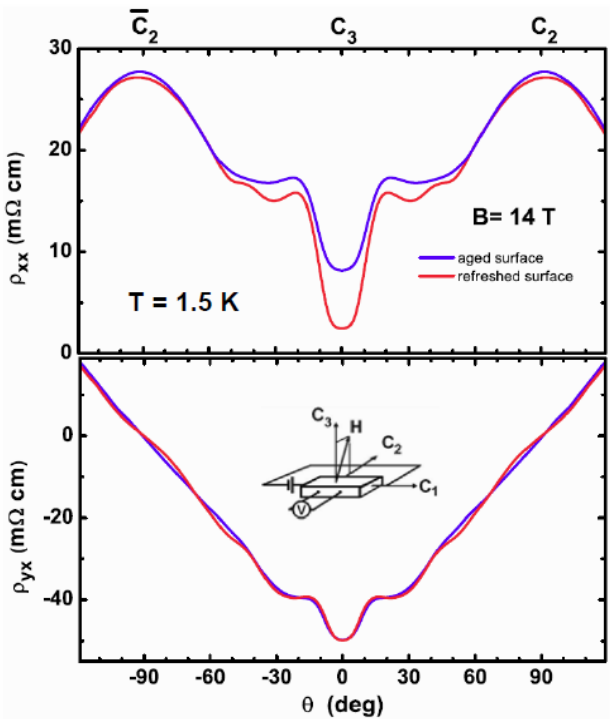

Figure 4. Direct comparison of the oscillations in $\rho_{xx}$ and $\rho_{yx}$ before and after etching the sample surface, measured on the same $Bi_{0.91}Sb_{0.09}$ sample. It is evident that the refreshed surface yields more pronounced oscillations, which gives evidence for the surface origin of these oscillations.

## 5. Conclusion

We present detailed data on the angular-dependent MR oscillation phenomenon which was recently discovered in a topological insulator $Bi_{0.91}Sb_{0.09}$. Direct comparison of the data taken before and after etching the sample surface gives compelling evidence that this novel phenomenon is essentially originating from a surface state. The symmetry of the oscillations suggests that it probably comes from the (111) plane. In the surface state of a topological insulator, there is no "quasi two-dimensionality" that introduces a finite warping to the 2D cylindrical Fermi surface, whereas the existence of such a warping is essential for the ordinary AMRO to occur; therefore, it is likely that a new mechanism, such as a coupling between the surface and the bulk states, is responsible for this intriguing phenomenon in topological insulators.

**Acknowledgments**
This work was supported by JSPS (KAKENHI Grant No. 19674002), MEXT (KAKENHI Grant No. 22103004), and AFOSR (AOARD Grant No. 10-4103).

**References**
[1] Qi X L and Zhang S C 2010 *Physics Today* **63** 33.
[2] Moore J E 2010 *Nature* **464** 194.
[3] Hasan M Z and Kane C L 2010 *arXiv:*1002.2895 (to be published in *Rev. Mod. Phys.*).
[4] Fu L. Kane C L and Mele E J 2007 *Phys. Rev. Lett.* **98** 106803.
[5] Fu L and Kane C L 2007 *Phys. Rev. B* **76** 045302.
[6] Moore J E and Balents L 2007 *Phys. Rev. B* **75** 121306(R).
[7] Qi X L, Hughes T L and Zhang S C 2008 *Phys. Rev. B* **78** 195424.
[8] Hsieh D *et al.* 2008 *Nature* **452** 970.
[9] Hsieh D *et al.* 2009 *Science* **323** 919.
[10] Nishide A *et al.* 2010 *Phys. Rev. B* **81** 041309(R).
[11] Eto K, Ren Z, Taskin A A, Segawa K and Ando Y 2010 *Phys. Rev. B* **81** 195309.
[12] Peng H *et al.* 2010 *Nature Materials* **9** 225.
[13] Checkelsky J G *et al.* 2009 *Phys. Rev. Lett.* **103** 246601.
[14] Taskin A A and Ando Y 2009 *Phys. Rev. B* **80** 085303.
[15] Taskin A A, Segawa K and Ando Y 2010 *Phys. Rev. B* **82** 121302(R).
[16] Kartsovnik M V *et al.* 1988 *JETP Lett.* **48** 541.
[17] Kajita K et al. 1989 *Solid St. Comm.* **70** 1189.
[18] Yamaji K 1989 *J. Phys. Soc. Jpn.* **58** 1520.
[19] Roushan P, Seo J, Parker C V, Hor Y S, Hsieh D, Qian D, Richardella A, Hasan M Z, Cava R J and Yazdani A 2009 *Nature* **460** 1106.
[20] Cooper B K and Yakovenko V M 2006 *Phys. Rev. Lett. 96* 037001.